# THE UNIFIED MODEL OF COMPACT RADIO SOURCES

Fedor V.Prigara

Institute of Microelectronics and Informatics, Russian Academy of Sciences, 21 Universitetskaya, 150007 Yaroslavl, Russia; fprigara@rambler.ru

## ABSTRACT

We propose the unified model of compact radio sources, i.e. of pulsars, maser sources, and active galactic nuclei. The unification is based on the wavelength dependence of radio source size. It is shown that the compact sources are characterized by a maser amplification of thermal radio emission. The density, temperature, and magnetic field profiles of compact sources are discussed.

*Subject headings: radiation mechanisms: thermal--AGN--pulsars--radio continuum*

## 1. INTRODUCTION

Compact radio sources have the small angular dimensions, usually less than 1 mas, and exhibit the high brightness temperatures. These properties are common for pulsars (Shklovsky 1984), masers (Bochkarev 1992), and active galactic nuclei (Bower & Backer 1998, Kellermann, Vermeulen, Zensus & Cohen 1998). The brightness temperatures of OH masers have the magnitude $T_b \leq 10^{12} K$, and those of water masers have the magnitude $T_b \leq 10^{15} K$ (Bochkarev 1992). Compact extragalactic sources (AGNs) exhibit brightness temperatures in the range of $10^{10}$ K to $10^{12}$ K (Bower & Backer 1998, Kellermann et al. 1998), so these temperatures have an order of magnitude of those of OH masers.

Another feature, which is common for the compact sources, is that their radio emission so far has not received a satisfactory explanation. In particular, it is true for pulsars (Qiao et al. 2000). In the case of maser sources the modern theory uses some chance coincidences which hardly can maintain in a more profound theory (Bochkarev 1992). At last, it was shown recently that spherical accretion models with the synchrotron mechanism of emission cannot explain the flat or slightly inverted radio spectra of low-luminosity active galactic nuclei (Nagar, Wilson & Falcke 2001, Ulvestad & Ho 2001).

## 2. THE GASEOUS DISK MODEL

It was shown recently (Prigara 2001b) that thermal radio emission has a stimulated character. According to this conception thermal radio emission of non-uniform gas is produced by an ensemble of individual emitters. Each of these emitters is a molecular resonator the size of which has an order of magnitude of mean free path $l$ of photons

$$l = \frac{1}{n\sigma} \tag{1}$$

where $n$ is the number density of particles and $\sigma$ is the photoabsorption cross-section.

The emission of each molecular resonator is coherent, with the wavelength

$$\lambda = l, \tag{2}$$

and thermal radio emission of gaseous layer is incoherent sum of radiation produced by individual emitters.

The condition (2) implies that the radiation with the wavelength $\lambda$ is produced by the gaseous layer with the definite number density of particles $n$.

In the gaseous disk model, describing radio emitting gas nebulae (Prigara 2001a), the number density of particles decreases reciprocally with respect to the distance $r$ from the energy center

$$n \propto r^{-1}. \tag{3}$$

Together with the condition of emission (2) the last equation leads to the wavelength dependence of radio source size:

$$r_\lambda \propto \lambda. \tag{4}$$

The relation (4) is indeed observed for sufficiently extended radio sources. For instance, the size of radio core of galaxy M31 is 3.5 arcmin at the frequency 408 MHz and 1 arcmin at the frequency 1407 MHz (Sharov 1982).

## 3. EXTENDED RADIO SOURCES

The spectral density of flux from an extended radio source is given by the formula

$$F_\nu = \frac{1}{a^2} \int_0^{r_\lambda} B_\nu(T) \times 2\pi r \, dr \quad , \tag{5}$$

where *a* is a distance from radio source to the detector of radiation, and the function $B_\nu(T)$ is given by the Rayleigh-Jeans formula

$$B_\nu = 2kT\nu^2/c^2, \qquad (6)$$

where ν is the frequency of radiation, *k* is the Boltzmann constant, and *T* is the temperature.

. The extended radio sources may be divided in two classes. Type 1 radio sources are characterized by a stationary convection in the gaseous disk with an approximately uniform distribution of the temperature *T≈const* giving the spectrum

$$F_\nu \approx const. \qquad (7)$$

Type 2 radio sources are characterized by outflows of gas with an approximately uniform distribution of gas pressure *P=nkT≈const*. In this case the equation (3) gives

$$T \propto r, \qquad (8)$$

so the radio spectrum, according to the equation (5), has the form

$$F_\nu \propto \nu^{-1}. \qquad (9)$$

Both classes include numerous galactic and extragalactic objects. In particular, edge-brightened supernova remnants (Kulkarni & Frail 1993) belong to the type 2 radio sources in accordance with the relation (8), whereas center-brightened supernova remnants belong to the type 1 radio sources.

## 4. THE WAVELENGTH DEPENDENCE OF RADIO SOURCE SIZE

In the case of compact radio sources instead of the relation (4) the relation

$$r_\lambda \propto \lambda^2 \qquad (10)$$

is observed (Lo et al. 1993, Lo 1982). This relation may be explained by the effect of strong magnetic field on the distribution of ionized gas density which changes the equation (3) for the equation

$$n \propto r^{-1/2}. \qquad (11)$$

In the limit of very strong magnetic fields the density of ionized gas does not depend on the radius *r*.

It is well known (Shklovsky 1984) that the delay of radio pulses from pulsars at low frequencies is proportional to $\lambda^2$. This fact is a mere consequence of Eq.(10), if we only assume the existence of the radial density wave traveling across the radius with a constant velocity and triggering the pulse radio emission. In this treatment the pulsars also obey the $\lambda^2$ dependence of compact source size. Note that the wavelength dependence of a pulse duration is a similar effect.

The spatial distribution of SiO, water, and OH masers (each of which emits in own wavelength) in the maser complexes also is consistent with the $\lambda^2$ dependence of compact source size (Bochkarev 1992, Eisner et al. 2001).

To summarize, extended radio sources are characterized by the relation (4), and compact radio sources obey the relation (10).

## 5. MASER AMPLIFICATION OF THERMAL RADIO EMISSION

The existence of maser sources associated with gas nebulae and galactic nuclei (Miyoshi et al. 1995) is closely connected with the stimulated origin of thermal radio emission. The induced origin of thermal radio emission follows from the relations between Einstein's coefficients for a spontaneous and induced emission of radiation (Prigara 2001b). The high brightness temperatures of compact, flat-spectrum radio sources (Bower & Backer 1998; Nagar, Wilson, & Falcke 2001; Ulvestad & Ho 2001) may be explained by a maser amplification of thermal radio emission. A maser mechanism of emission is supported by a rapid variability of total and polarized flux density on timescales less than 2 months (Bower et al. 2001). Such a variability is characteristic for non-saturated maser sources. Note that the spherical accretion models with the synchrotron mechanism of emission are unable to explain the flat or slightly inverted spectra of low-luminosity active galactic nuclei (Nagar et al. 2001; Ulvestad & Ho 2001). The Blandford and Konigl theory used by Nagar et al. (2001) is in some respects similar to the gaseous disk model (Prigara 2001a), the latter being more simple and free from indefinite parameters, such as an empirical spectral index.

It is shown by Siodmiak & Tylenda (2001) that the standard theory of thermal radio emission which does not take into account the induced character of emission cannot explain the radio spectra of planetary nebulae at high frequencies without an introduction of indefinite parameters.

## 6. THE ZEEMAN SPLITTING OF RADIO PULSES FROM PULSARS

The observed radio profiles of pulsars have several components, one to five (Qiao et al. 2000). This splitting of radio pulses can be explained by the Zeeman splitting of radio lines in the magnetic field of a pulsar together with a spatial separation of the layers emitting in various frequencies.

The spatial separation of the layers emitting in various frequencies is produced by the condition for emission (2), which is likely to be valid in the case of radio emission of pulsars, as well as in the case of thermal radio emission. The reason is that both kinds of emission have an induced origin. A maser mechanism of radiation process in pulsars is supported by 1) the high brightness temperatures; 2) the rapid variability of the pulse profiles (Shklovsky 1984); 3) a pulse character of emission; 4) a strong linear and circular polarization of emission.

The last is produced by the Zeeman splitting of radio lines. A pulse character of emission and a spatial separation of the emitting layers transform the frequency splitting into the time splitting of the pulse at a given frequency. It is assumed that the number density of particles is decreasing with the increase of the distance from the energy center of a pulsar in accordance with the gaseous disk model.

In the case of free electrons, the Zeeman splitting in the magnetic field $B$ is $h\Delta\nu = 2\mu B$, where $h$ is the Planck constant, and $\mu$ is the Bohr magneton. If $B=1$ Gs, then $\Delta\nu=3$ MHz. At a given frequency $\nu$ two pulses are observed from the emitting layers corresponding to the non-shifted frequencies $\nu - \frac{1}{2}\Delta\nu$ and $\nu + \frac{1}{2}\Delta\nu$. This is likely to be the case of type II c pulsars, according to the classification by Qiao et al. (2000). The pulse profiles of this type pulsars are double at all frequencies, with the separation between two peaks decreasing at higher frequencies.

According to the unified model of compact radio sources the radius $r_\nu$ of the layer emitting in the frequency $\nu$ is proportional to $\nu^{-2}$. If the magnetic field profile has a form $B \propto r^{-\alpha}$, then the Zeeman splitting obeys the law $\Delta\nu \propto B \propto r_\nu^{-\alpha} \propto \nu^{2\alpha}$.

The separation between the two components of the pulse is given by the formula

$$\tau = \Delta r_\nu / v \propto \Delta\nu / \nu^3 \propto \nu^{2\alpha-3} \qquad (12)$$

where $v$ is the velocity of the radial density wave triggering the emission process.

Since the time separation between the two peaks decreases at higher frequencies, for the power law index $\alpha$ we have an estimation $\alpha < 3/2$.

Note that the magnetic field cannot be uniform in the emitting region, since the direction of linear polarization changes within the pulse profile.

## 7.CONCLUSIONS

The compact radio sources are characterized by the following properties: 1) the small angular dimensions; 2) the high brightness temperatures; 3) the $\lambda^2$ dependence of radio source size; 4) the maser mechanism of radio emission.

Exactly by the maser mechanism of emission the compact radio sources differ from the extended ones, from the physical point of view.

The properties of the compact radio sources can be adequately described within the gaseous disk model with a suitable density profile. The unified model of compact radio sources gives, in particular, a natural explanation of the splitting of radio pulses from pulsars.